# Blind Turing-Machines: Arbitrary Private Computations from Group Homomorphic Encryption

Stefan Rass

System Security Group, Institute of Applied Informatics
Alpen-Adria Universität Klagenfurt
Klagenfurt, Austria

*Abstract*—Secure function evaluation (SFE) is the process of computing a function (or running an algorithm) on some data, while keeping the input, output and intermediate results hidden from the environment in which the function is evaluated. This can be done using fully homomorphic encryption, Yao's garbled circuits or secure multiparty computation. Applications are manifold, most prominently the outsourcing of computations to cloud service providers, where data is to be manipulated and processed in full confidentiality. Today, one of the most intensively studied solutions to SFE is fully homomorphic encryption (FHE). Ever since the first such systems have been discovered in 2009, and despite much progress, FHE still remains inefficient and difficult to implement practically. Similar concerns apply to garbled circuits and (generic) multiparty computation protocols. In this work, we introduce the concept of a blind Turing-machine, which uses simple homomorphic encryption (an extension of ElGamal encryption) to process ciphertexts in the way as standard Turing-machines do, thus achieving computability of any function in total privacy. Remarkably, this shows that fully homomorphic encryption is indeed an overly strong primitive to do SFE, as group homomorphic encryption with equality check is already sufficient. Moreover, the technique is easy to implement and perhaps opens the door to efficient private computations on nowadays computing machinery, requiring only simple changes to well-established computer architectures.

## I. INTRODUCTION

Many security systems at some point employ trusted parties (e.g., trust-centers, smartcards) to perform computations on secret (confidential) information. Trying to get rid of such trusted instances in a security system is often difficult (if not impossible), and one possible solution is to emulate the trusted party by a collection of instances rather than a single one. Such distributed computations usually rely on secret-sharing techniques, capable of function evaluation such that only a permitted specified set of coalitions can learn any secret information or results of the computation. The work of Gennaro[1][2] and [3]made significant contributions to the theory in this area known as *secure multiparty computation* (SMC). Its practical usefulness, however, is somewhat limited, as it isoften tied to a vast communication effort and intricate additional security precautions (e.g., pairwise confidential channels, broadcast channels, etc.). Furthermore, it is a special case ofthe more general problem known as *secure function evaluation* (SFE), in which a single (potentially malicious) instance is made to compute some function on externally supplied (potentially encrypted) inputs. This is the area where this work falls into, and on which we will exclusively concentrate us in the following.Commoncomputational models upon which SFE is based are Turing-machines or circuits, where the appropriateness of each model depends on the details of the SFE technique. We will base our construction on Turing-machines, drawing strongly from circuit complexity models to ease life in cryptographic matters.

**Related work**: One famous approach to SFE, leaving the computations with a single not necessarily trusted entity,is provided by Yao's *garbled circuits* (GC) [4]. Here, the computational model are circuits, which are good for hardware-implementation (as well as theoretical treatment), yet somewhat difficult to apply in a generic fashion to handle inputs of arbitrary size. Despite much progress in this direction [5]-[7] as well as on applications of GC for SMC [8], [9], only uniform circuits can be set up effectively in practice, in which case they are essentially equivalent to Turing-machines. However, there is so far no analogous concept of a garbled Turing-machine.

Without doubt, the most powerful (and recent) solution to SFE is fully homomorphic encryption (FHE). In brief, this is (or can be) a trapdoor one-way automorphism $E: (R, +, \cdot) \times \mathcal{K} \to (R, +, \cdot)$ where $R$ is a ring or a field, and $\mathcal{K}$ is the key-space. We denote the encryption $E(m, k)$ of $m$ under the key $k$ as $E_k(m)$ hereafter. The central property of FHE is its compatibility with arithmetic operations in the sense that for any two plaintexts $m_1, m_2$ and any key $k$, we get $E_k(m_1) + E_k(m_2) = E_k(m_1 + m_2)$ and $E_k(m_1 \cdot m_2) = E_k(m_1) \cdot E_k(m_2)$. That is, arithmetic manipulations done to ciphertexts identically apply to the underlying plaintexts. It is easy to imagine that such an encryption enables any kind of data processing given ciphertexts only, which is exactly what secure function evaluation means.While it is usually simple to get a homomorphic property w.r.t. addition *or* multiplication (e.g., standard encryptions such as RSA or ElGamal are multiplicatively homomorphic; in general *group homomorphic*), homomorphy w.r.t. both operations is intricate and has only recently been achieved [10]. Since this breakthrough, FHE has evolved into a major research branch of cryptography, with many interesting results [11]-[18].

**Our contribution** in this work is to show that despite the theoretical beauty of FHE, it is nevertheless an overly strong primitive for secure function evaluation. To this end, we investigate the weaker notion of *public-key encryption with equality check* (PKEET) [19], and show how the functionality of a basic (single two-way infinite tape) Turing-machine can be implemented with simple homomorphic encryption that allows equality checks. We call the resulting computing model a *blind Turing-machine*, since it works on encrypted tape content only, doing its transitions by virtue of equality checks, and manipulating the tape content using the homomorphic properties of the encryption in charge. Hence, the TM does not see (in plaintext) any of the content that it processes.

The rest of this article is organized as follows: we start from the PKEET system of [19], which is secure under adaptive chosen ciphertext attacks, and as such cannot be in any sense homomorphic. To restore the homomorphic property in the framework of [19], we describe a generic technique (based on [19]) to construct a *homomorphic public-key encryption with equality check* (hereafter abbreviated as HPKEET) from any additively homomorphic encryption. We prove it secure under (non-adaptive) chosen ciphertext attacks (stronger notions are provably unachievable for any homomorphic encryption in general). Section II formally introduces the respective models, with the construction of HPKEET and its security analysis to follow in section III. Blind Turing-machines (BTM) are introduced in section IV, based on a brief review of how conventional Turing-machines (TM) are formally defined. Security and complexity of computations on such blind TM are studied along a sequence of subsections.

In section VII we report on a practical implementation of the encryption. Remarks on future work and open problems follow in section VIII.

## II. DEFINITIONS

We write $x \xleftarrow{r} X$ to denote a uniformly random draw of an element $x$ from a set $X$. We let $|x|$ denote the length of $x$ in bits (assuming a canonical string representation, if $x$ is a group element). Our treatment in the following is non-uniform. That is, we consider the *complexity* of an algorithm as the *size*, i.e. the number of gates, of a circuit representing the algorithm. To handle inputs of varying length, a *circuit* (e.g., adversary) of complexity $\tau$ is thus to be understood as a sequence of circuits (circuit family) $(C_n)_{n \in \mathbb{N}}$, whose size is a function $\tau(n)$, whenever the circuit $C_n$ has $n$ input gates. Besides circuit complexity, section V will heavily rest on time-complexity considerations. To distinguish the two notions from each other, we will refer to *circuit complexity* simply as *complexity*, as opposed to *time-complexity*, always carrying the prefix „time". To further clarify which concept is in charge, we will speak of *circuits* to mean circuit families and circuit complexity, and *algorithms* when we consider time-complexities.

A public-key encryption scheme is a triple of circuits $(G, E, D)$. The circuit $G$ generates the secret and public key pair, denoted as $(pk, sk) \leftarrow G$. For homomorphy, assume that the encryption function $E: \mathbb{G}_m \times \mathcal{K} \to \mathbb{G}_c$ is defined on a cyclic plaintext group $(\mathbb{G}_m, +)$, keyspace $\mathcal{K} \subseteq \{0,1\}^*$ and cyclic ciphertext group $(\mathbb{G}_c, \cdot)$. Abbreviating the encryption of a plaintext $m$ under the public key $pk$ by $E_{pk}(m)$, we require group homomorphy under identical public keys, i.e., $E_{pk}(m_1) \cdot E_{pk}(m_2) = E_{pk}(m_1 + m_2)$ for all $m_1, m_2 \in \mathbb{G}_m$. The function $D: \mathbb{G}_c \times \mathcal{K} \to \mathbb{G}_m$ decrypts a ciphertext $c$ upon given the secret key $sk$; denoted as $D_{sk}(c)$.

Security of an encryption is commonly defined in terms of indistinguishable ciphertexts under differently strong attack scenarios. However, an indistinguishability requirement is obviously useless once we endow an encryption with comparison facilities for plaintexts that work on ciphertexts only (as we attempt here). To fix this, we additionally introduce an *authorization* function (circuit) that outputs a (secret) *comparison key*, hereafter called a *token*, which enables comparisons, while any party not knowing the token will be unable to distinguish any two given ciphertexts. In that sense, we consider two different kinds of attacker (following the framework of [19]), both of which are given all system parameters and public keys:

**Type 1 attacker**: This one can do ciphertext comparisons, in which case we can only ensure the cipher to be one-way but not indistinguishable.

**Type 2 attacker**: This one does *not* have the authorization token to do comparisons, thus security against this (weaker) attacker can properly be defined in terms of indistinguishability.

Onewayness under chosen-ciphertext attacks is defined in the usual way by giving oracle access to $D_{sk}(\cdot)$ to the attacker $\mathcal{A}$, indicated as $\mathcal{A}^{D_{sk}(\cdot)}$, and engaging in the following experiment $\mathbf{Exp}^{\mathcal{A}}_{\text{OW-CCA1}}$ with the challenger.

**Setup phase**: the challenger creates $(pk, sk) \leftarrow G$.

**Query phase**: the attacker (adaptively) chooses a number of $q$ ciphertexts $c_i \in \mathbb{G}_c$ and retrieves $m_i \leftarrow D_{sk}(c)$ from the challenger for $i = 1, 2, \ldots, q$.

**Challenge phase**: challenger chooses a plaintext $m$ that has not been returned in the query phase, and submits $c^* \leftarrow E_{pk}(m)$ to the adversary.

**Guess phase**: attacker outputs a guess $m^*$.

The *advantage* of $\mathcal{A}^{D_{sk}(\cdot)}$ in $\mathbf{Exp}_{\text{OW-CCA1}}$ is

$$\mathbf{Adv}^{\mathcal{A}^{D_{sk}(\cdot)}}_{\text{OW-CCA1}} := \Pr[D_{sk}(c^*) = m^* | m^* \leftarrow \mathcal{A}^{D_{sk}(\cdot)}].$$

We call the encryption $(\tau, q, \varepsilon)$-*OW-CCA1-secure*, if an adversary $\mathcal{A}^{D_{sk}(\cdot)}$ of complexity $\leq \tau$ and submitting no more than $q$ queries has an advantage $\mathbf{Adv}^{\mathcal{A}^{D_{sk}(\cdot)}}_{\text{OW-CCA1}} < \varepsilon$.

Indistinguishability under chosen-ciphertext attacks is defined by the following experiment $\mathbf{Exp}^{\mathcal{A}}_{\text{IND-CCA1}}$. As before, we assume oracle access to decryptions under $sk$:

**Setup phase**: the challenger creates $(pk, sk) \leftarrow G$.

**Query phase**: the attacker (adaptively) chooses a number of $q$ ciphertexts $c_i \in \mathbb{G}_c$ and retrieves $m_i \leftarrow D_{sk}(c)$ from the challenger for $i = 1, 2, \ldots, q$.

**Challenge phase**: the attacker generates two messages $m_0, m_1 \leftarrow \mathcal{A}^{D_{sk}(\cdot)}$, where $m_0 \neq m_1$ and $|m_0| = |m_1|$. The challenger receivers $m_0, m_1$, chooses $b \xleftarrow{r} \{0,1\}$, and returns $c_b \leftarrow E_{pk}(m_b)$.

**Guess phase**: attacker outputs a guess $b^*$.

The *advantage* of $\mathcal{A}^{D_{sk}(\cdot)}$ in $\mathbf{Exp}_{\text{IND-CCA1}}$ is

$$\mathbf{Adv}_{\text{IND-CCA1}}^{\mathcal{A}^{D_{sk}(\cdot)}} := \left| \Pr[b^* = b] - \frac{1}{2} \right|.$$

We call the encryption $(\tau, q, \varepsilon)$ *-IND-CCA1-secure*, if an adversary $\mathcal{A}^{D_{sk}(\cdot)}$ of complexity $\leq \tau$ and submitting no more than $q$ queries has an advantage $\mathbf{Adv}_{\text{IND-CCA1}}^{\mathcal{A}^{D_{sk}(\cdot)}} < \varepsilon$.

Comparisons can be done by allowing decryptions of either the plaintext or a hash-value thereof (not revealing the plaintext as such). To this end, we define a commitment-like hash-function that acts on the same plaintext group as the encryption does. For security, we require the discrete logarithm problem to be difficult on this group. Formally, let $g \in \mathbb{G}_c$ generate the group $\mathbb{G}_c$. We call $\mathbb{G}_c$ a $(\tau, \varepsilon)$ -DL group, if $\Pr\left[ x = x' \middle| x \xleftarrow{r} \mathbb{G}_m, y = g^x, x' \leftarrow \mathcal{A}(y) \right] < \varepsilon$ for all circuits $\mathcal{A}$ of complexity $\leq \tau$.

### A. Asymptotic Security

For generality, we give concrete security statements here in terms of the security parameters $\tau, q, \varepsilon$, leaving their obvious respective asymptotic formulations aside. Throughout the rest of this work, we confine ourselves to stressing that all parameters, in the asymptotic formulation, would depend on a (common) security parameter $\kappa \in \mathbb{N}$ that usually controls key-sizes, group structures or similar (consequently, it goes as a parameter into the key-generation circuit $G$). As an example, the asymptotic version of $(\tau, q, \varepsilon)$-OW-CCA1 security would read as follows: for every polynomials $\tau(\kappa), q(\kappa)$ there is a negligible function [1] $\varepsilon(\kappa)$ such that the encryption is $(\tau(\kappa), q(\kappa), \varepsilon(\kappa))$ -OW-CCA1 secure. All results and definitions to follow can be restated in a similar manner.

### III. THE ENCRYPTION SCHEME

Our encryption scheme will allow comparisons by attaching a keyed hash of the inner plaintext to the ciphertext, where the key for the hash is also encrypted [2]. Comparisons then need the permission by the originator of the ciphertext, who must provide the decryption key to disclose the hash-key. This key is obtained by an *authorization function* Aut. The comparison procedure com then simply compares the "decrypted" hashes. To distinguish the components of our HPKEET-encryption scheme (KeyGen, Enc, Dec, Aut, Com) from that of the underlying OW-CCA1 and IND-CCA1 secure encryption $(G, E, D)$, we use a different notation hereafter. Moreover, we assume that the plaintext group $\mathbb{G}_m$ is such that DL-commitments to $m \in \mathbb{G}_m$ are well-defined; that is, we can compute $g^m$ for a generator $g$ of $\mathbb{G}_c$ and some value $m \in \mathbb{G}_m$. This is trivially satisfied for prime order groups over the integers (say, if $\mathbb{G}_c \simeq \mathbb{G}_m \simeq \mathbb{Z}_p$ for some prime $p$ ), and requires only simple additional measures in elliptic curve settings.

KeyGen: Create $(pk_1, sk_1) \leftarrow G$ and $(pk_2, sk_2) \leftarrow G$. Put $pk := (pk_1, pk_2)$ and $sk := (sk_1, sk_2)$. Choose two (distinct) generators $g, h$ of $\mathbb{G}_c$. The system parameters globally known to all instances are $\mathbb{G}_m, \mathbb{G}_c, g$ and $h$.

Enc: Given the message $m \in \mathbb{Z}$ , the encryption is Enc: $\mathbb{G}_m \rightarrow \mathbb{G}_c^3$ by choosing an integer $r < |\mathbb{G}_c|$ and returning $\text{Enc}_{pk}(m) := (E_{pk_1}(m), g^m h^r, E_{pk_2}(r))$ .

Dec: If the given ciphertext $c$ cannot be parsed as an element $(c_1, c_2, c_3) \in \mathbb{G}_c^3$ , return $\perp$ . Otherwise, put $m' \leftarrow D_{sk_1}(c_1), r' \leftarrow D_{sk_2}(c_3)$ and verify if $c_2 = g^{m'} h^{r'}$. Output $m'$ upon a match, and $\perp$ otherwise.

Aut: To authorize a third party to do comparisons, Aut extracts and returns the *token* $t \leftarrow sk_2$ from the secret key $sk = (sk_1, sk_2)$.

Com: Given a token $t$ and two (syntactically correct) ciphertexts $(c_1, c_2, c_3), (c_1', c_2', c_3') \in \mathbb{G}_c^3$ , compute $r \leftarrow D_t(c_3), r' \leftarrow D_t(c_3')$ and output the result of the comparison $c_2 \cdot h^{-r} = c_2' \cdot h^{-r'}$ in $\mathbb{G}_c$.

Notice that a trivial instantiation of the above scheme by a symmetric (e.g., AES) or deterministic (e.g. plain RSA) encryption would be insecure. Even though comparisons are easy in that case (ciphertext equality implies plaintext equality), such a scheme would not be indistinguishable, and thus fail to achieve the security that we desire against a type 2 attacker.

### A. Homomorphy

Let two ciphertexts $c_i = (E_{pk}(m_i), g^{m_i} h^{r_i}, E_{pk}(r_i))$ for $i = 1, 2$ be given, and consider their component-wise product in $(\mathbb{G}_c, \cdot)$, which is

$$\begin{aligned} c_1 \cdot c_2 &= [E_{pk}(m_1) \cdot E_{pk}(m_2), g^{m_1} h^{r_1} g^{m_2} h^{r_2}, \\ &\qquad E_{pk}(r_1) \cdot E_{pk}(r_2)] \\ &= (E_{pk}(m_1 + m_2), g^{m_1 + m_2} h^{r_1 + r_2}, E_{pk}(r_1 + r_2)) \end{aligned}$$

This is a valid ciphertext if and only if the underlying encryption $(G, E, D)$ is *additively homomorphic*. Unfortunately, we cannot instantiate $(G, E, D)$ as a Paillier-encryption, since this works over a composite modulus $n = pq$ for which $\mathbb{Z}_n$ is not cyclic (in general). An "almost" compatible IND-CCA1 secure encryption, except for its multiplicative homomorphy, is found in Damgårds version of ElGamal encryption [20]. Changing the multiplicative homomorphic property of ElGamal encryption into an additive one is easy by encrypting commitments $g^m$ instead of $m$, if the plaintext space is only of "tractably small size" (e.g., polynomial size in the security parameter) to let us recover $m$ from $g^m$ efficiently. While this requirement is easily met in our application to Turing-machines, we stress that care has to be taken in the encoding of $m$ in order to avoid trial opening of commitments (and thus breaking the encryption) during an invocation of Com, if the token (secret key to decrypt the randomizer) is available (through an invocation of Aut). We take a closer look at this now.

---

[1] Negligibility of a function $f$ is defined in the usual way, by requiring for every $\alpha > 0$ the existence of a constant $K_\alpha > 0$ so that $f(\kappa) < \frac{1}{\kappa^\alpha}$ as soon as $\kappa > K_\alpha$.

[2] This proposal is as well found in [19], where it is instantiated in an insecure manner under an adaptive chosen ciphertext attack scenario (CCA2). We consider a similar instantiation (equally well not IND-CCA2 secure), but prove it secure in the weaker model of (non-adaptive) chosen-ciphertext attacks (IND-CCA1).

### B. Security Analysis

We start with a (well-known) necessary condition for security to avoid brute-force plaintext search.

#### 1) Offline Message Recovery

Plaintext discovery by trial encryptions and checking equality with the given ciphertext is essentially unavoidable, but can be made infeasible if the plaintexts have high min-entropy: recall that a random plaintext $M$ over a set $\mathbb{G}_m$ has min-entropy $H_\infty(M) = k$, if $k$ is the largest number such that $\Pr[M = m] \leq 2^{-k}$ for all $m \in \mathbb{G}_m$.

**Lemma 1.** If an encryption function $E$ is such that for any circuit $A$ of complexity $\tau$, we have $\Pr[M = m^* | m^* \leftarrow A(c)] < \varepsilon$ for any given ciphertext $c = E_{pk}(M)$, then the plaintext $M$ has min-entropy

$$H_\infty(M) > \log_2 \left[ 1 - (1 - \varepsilon)^{t_{ED} + \frac{t_{comp}}{\tau}} \right], \qquad (1)$$

Where $t_{ED}$ is the complexity of computing an encryption, and $t_{comp}$ measures how much circuitry is required to string-compare two ciphertexts.

*Proof.* If the lemma were wrong, then a circuit can do encryptions (of complexity $t_{ED}$) and comparisons (of complexity $t_{comp}$) to determine the correct plaintext. From the geometric distribution, it is easy to obtain the number of trials until the success probability becomes $\geq \varepsilon$. Constraining this number to be less than $\tau/(t_{ED} + t_{comp})$ (assuming the circuitry to be divided equally into blocks that do encryptions and comparisons), gives the stated min-entropy bound. □

Lemma 1 is a necessary yet insufficient condition for security. Its asymptotic counterpart (i.e., $H_\infty(M) \in \omega(\kappa)$ when $\kappa$ is the security parameter) is a standard requirement for security of deterministic or searchable encryption (cf. [21]) against polynomial time-bounded attackers. We establish security of the encryption as such in the next section, and postpone a discussion on how to practically assure condition (1) until section IV.B.

#### 2) Chosen Ciphertext Security

As the encryption comes with comparison facilities, we modify the OW-CCA1 and IND-CCA1 games appropriately, by letting the attacker submit Aut-queries besides decryption requests. To distinguish the experiments concerning HPKEET from that on the underlying cipher $(G, E, D)$, we denote these extended versions as $\mathbf{Exp}^A_{\text{OW-CCAE1}}$ and $\mathbf{Exp}^A_{\text{IND-CCAE1}}$, i.e., security under chosen ciphertexts and equality checks. The definition of $\mathbf{Exp}^A_{\text{OW-CCAE1}}$ is the same as that of $\mathbf{Exp}^A_{\text{OW-CCA1}}$, except for a slight modification in the query phase:

$\mathbf{Exp}_{\text{OW-CCAE1}}$ **query phase**: the attacker submits no more than $q$ queries of the giving $m_i \leftarrow D_{sk}(c_i)$ for (adaptively) chosen ciphertexts $c_i$ or $t \leftarrow$ Aut, for an authorization query.

Obviously, we cannot apply the same change to $\mathbf{Exp}_{\text{IND-CCAE1}}$, so we define this experiment *exactly identical* to $\mathbf{Exp}_{\text{IND-CCA1}}$.

By construction, our encryption is a humble application of $E$ on two stochastically independent quantities $m$ and $r$, along with a product of two commitments thereof. Hence, the reductions establish only a slight advantage over that in breaking $(G, E, D)$. Formally, we have

**Lemma 2.** Let $(G, E, D)$ be defined over an $(\tau + t_{ED} + t_{eim}, \varepsilon_{DL})$-DL-group $\mathbb{G}_m$ of plaintexts, where $t_{ED}$ is the maximum complexity of an encryption or decryption, and $t_{eim}$ is the total complexity of one exponentiation with inversion and multiplication in $\mathbb{G}_c$. If $(G, E, D)$ is $(\tau + (q + 1)(t_{ED} + t_{eim}), q, \varepsilon)$ -OW-CCA1-secure, then the corresponding HPKEET scheme is $(\tau, q, \varepsilon + \varepsilon_{DL})$ -OW-CCA1-secure.

*Proof.* Suppose the existence of an attacker $A$ with advantage $\mathbf{Adv}^{A^{\text{Dec}_{sk}(\cdot)}}_{\text{OW-CCAE1}} > \varepsilon + \varepsilon_{DL}$ and complexity $\leq \tau$ and making $q$ queries. We construct an attacker $A'$ that wins $\mathbf{Exp}^{A'}_{\text{OW-CCA1}}$ as follows: given $(pk, sk)$ from $\mathbf{Exp}_{\text{OW-CCA1}}$, $A'$ sets $(pk_1, sk_1) := (pk, sk)$ and obtains $(pk_2, sk_2) \leftarrow G$ on its own. It then simulates $\mathbf{Exp}^A_{\text{OW-CCAE1}}$ for $A$, answering the $i$-th query (for $i = 1, 2, \ldots, q$) as follows:

- Dec-queries on an incoming HPKEET ciphertext $c_i = (c_{i,1}, c_{i,2}, c_{i,3})$ are forwarded as decryption challenges $c_i = c_{i,1}$ to the OW-CCA1 challenger, which returns $m' \leftarrow \text{Dec}_{sk}(c_i) = D_{sk_1}(c_{i,1})$. Then, $A'$ goes on by decrypting $c_{i,3}$ using its own secret key $sk_2$ into $r' \leftarrow \text{Dec}_{sk_2}(c_{i,3})$, and returns $m'$ if $g^{m'} h^{r'} = c_{i,2}$ in $\mathbb{G}_c$, and $\bot$ otherwise. We stress that the keypairs $(pk_1, sk_1)$ and $(pk_2, sk_2)$ for encrypting the payload and the randomizer are in any case chosen stochastically independent. Hence, $A'$ actually acts properly if it generates $(pk_2, sk_2)$ by itself, and receives the other pair $(pk_1, sk_1)$ from an external source (the OW-CCA1 challenger).

- Aut-queries are answered faithfully by responding with $sk_2$.

In the challenge phase, the complexity of $A$ is thus dominated by simulations of half of the decryption of challenges from $A$, which is $\leq q \cdot (t_{ED} + t_{eim})$.

To ease notation, let us incorporate all information from the query phase of $\mathbf{Exp}^A_{\text{OW-CCAE1}}$ into the circuit $A$, which in the guess phase of $\mathbf{Exp}^A_{\text{OW-CCAE1}}$ computes its output upon a given ciphertext $(c_1^*, c_2^*, c_3^*) = (E_{pk_1}(m), g^m h^r, E_{pk_2}(r))$. Observe that $c_3^*$, in an information-theoretic sense, does not provide any information on $c_1^*$, and uniquely determines $g^m$ from $c_2^*$. Therefore, in any $\mathbf{Exp}^A_{\text{OW-CCAE1}}$ execution in which at least one Aut-query has been submitted,

$$\mathbf{Adv}^{A^{\text{Dec}_{sk}(\cdot)}}_{\text{OW-CCAE1}}$$
$$= \Pr[\text{Dec}_{sk_1}(c^*) = m^* | m^* \leftarrow A(E_{pk_1}(m), g^m h^r, E_{pk_2}(r))]$$
$$= \Pr[\text{Dec}_{sk_1}(c^*) = m^* | m^* \leftarrow \hat{A}(E_{pk_1}(m), g^m)]$$

for some circuit $\hat{\mathcal{A}}$. Obviously, one could convert from the inputs $(E_{pk_1}(m), g^m)$ to $(E_{pk_1}(m), g^m h^r, E_{pk_2}(r))$ (and back) by choosing (or decrypting) the randomizer $r$ and doing (or inverting) the remaining operations. Hence, the complexity of $\hat{\mathcal{A}}$ is bounded from above by $\tau + t_{ED} + t_{eim}$, where $t_{ED}$ and $t_{eim}$ are the complexities of an encryption/decryption (maximum thereof), and an exponentiation with inversion and multiplication in $\mathbb{G}_c$. The advantage of $\hat{\mathcal{A}}$ is the probability of guessing $m^*$ correctly either from $E_{pk_1}(m)$ or $g^m$ alone, or from both. From the union bound and by assuming that $\mathbb{G}_c$ is an $(\tau + t_{ED} + t_{eim}, \varepsilon_{DL})$-DL-group, we get

$$
\begin{aligned}
\varepsilon + \varepsilon_{DL} \quad < \quad & \mathbf{Adv}^{\hat{\mathcal{A}}}_{\text{OW-CCAE1}} \\
\leq \quad & \Pr[m = m^* | m^* \text{ extracted from} E_{pk_1}(m)] \\
& + \Pr[m = m^* | m^* \text{ extracted from} g^m] \\
= \quad & \mathbf{Adv}^{\mathcal{A}'}_{\text{OW-CCA1}} + \varepsilon_{DL}.
\end{aligned}
$$

The complexity of $\mathcal{A}'$ is thus $q \cdot (t_{ED} + t_{eim}) + \tau + t_{ED} + t_{eim} = \tau + (q+1)(t_{ED} + t_{eim})$.

In the challenge phase of $\mathbf{Exp}^{\mathcal{A}'}_{\text{OW-CCA1}}$, upon incoming of $c^* = E_{pk}(m)$, $\mathcal{A}'$ can therefore run $\hat{\mathcal{A}}$ in place of $\mathcal{A}$, to discover $m$ from the OW-CCA1 challenge $c^* = E_{pk}(m)$, with an advantage $\mathbf{Adv}^{\mathcal{A}'}_{\text{OW-CCA1}} > \varepsilon$ contradicting the security of $(G, E, D)$. □

Likewise, we establish IND-CCA1-security of HPKEET by virtue of the following well-known concrete result on how indistinguishability implies semantic security.

**Lemma 3.** If $(G, E, D)$ has $(\tau, \varepsilon)$-indistinguishable encryptions, and $E_{pk}$ has complexity $\leq t_{ED}$, then $(G, E, D)$ is $(\tau - \ell_f, t_{ED}, \varepsilon)$-semantically secure where $(t_1, t_2, \delta)$-semantic security is defined as follows: for every distribution $X$ over messages, every functions $I: \mathbb{G}_m \to \{0,1\}^*$, $F: \mathbb{G}_m \to \{0,1\}^{\ell_f}$ (of arbitrary complexity) and every circuit $A$ of complexity $\leq t_1$, there is another circuit $A^*$ with complexity $\leq t_1 + t_2$ so that

$$
\begin{aligned}
\Big| \Pr\Big[ A\Big(E_{pk}(m), I(m)\Big) = F(m) \Big] \\
- \Pr[A^*(I(m)) = F(m)] \Big| < \varepsilon.
\end{aligned}
$$

**Lemma 4.** Let $(G, E, D)$ be defined over a group $\mathbb{G}_m$ of plaintexts, where $t_{ED}$ bounds the complexity of an encryption or decryption, and $t_{eim}$ is the complexity of one exponentiation with inversion and multiplication in $\mathbb{G}_c$. If $(G, E, D)$ is $(\tau + q \cdot (t_{ED} + t_{eim}) + t_{ED} - 1, q, \varepsilon)$-IND-CCA1-secure, then the corresponding HPKEET scheme is $(\tau, q, 2\varepsilon)$-IND-CCA1-secure.

*Proof.* Besides a few modifications that we describe now, the line of arguments is completely analogous as in the proof of Lemma 2, except for the important difference that the adversary is not allowed to issue $\mathtt{Aut}$-queries in $\mathbf{Exp}^{\mathcal{A}}_{\text{IND-CCAE1}}$.

Assume an attacker $\mathcal{A}$ with $2\varepsilon$-advantage in $\mathbf{Adv}^{\mathcal{A}}_{\text{IND-CCAE1}}$. The complexity of $\mathcal{A}$ during the challenge phase is (as before) $q \cdot (t_{ED} + t_{eim})$. Upon the incoming challenge $c_b = E_{pk}(m_b)$ in $\mathbf{Exp}^{\mathcal{A}}_{\text{IND-CCA1}}$, $\mathcal{A}$ embeds it in a HPKEET ciphertext

$c_b^* = (c_b, r, r')$, for $r, r' \xleftarrow{r} \mathbb{G}_m$. Observe that a unique value $r'' \in \mathbb{G}_m$ exists for which $g^m h^{r''} = r$. For $c_b^*$ to be a valid HPKEET ciphertext, $r'$ should equal $E_{pk_2}(r')$, which is most likely not the case. We can fix this by exploiting the indistinguishability of encryptions under $E$ as follows: as $E$ is $(\tau + q \cdot (t_{ED} + t_{eim}), q, \varepsilon)$-IND-CCA1-secure, Lemma 3 lets us replace $\mathcal{A}$ by another circuit $\mathcal{A}^*$ that has complexity $\tau + q \cdot (t_{ED} + t_{eim}) + t_{ED} - 1$ and delivers the decision $f(c_b) = b' \in \{0,1\}^{\ell_f = 1}$ so that

$$
\begin{aligned}
\Big| \Pr\Big[ \mathcal{A}\Big(E_{pk}(m_b), g^m h^{r'}, E_{pk}(r')\Big) = b \Big] \\
- \Pr[\mathcal{A}^*(E_{pk}(m_b), g^m h^{r'}) = b] \Big| \leq \varepsilon
\end{aligned} \tag{2}
$$

Observe that $g^m h^{r'} = g^{m_0} h^{r''}$ for some random $r''$, which means that this second parameter to $\mathcal{A}^*$ – in an information-theoretic sense – does not provide any additional information on $b$. So, there is another circuit $\mathcal{A}^{**}$, no more complex than $\mathcal{A}^*$, such that $\Pr[\mathcal{A}^*(E_{pk}(m_b), g^m h^{r'}) = b] = \Pr[\mathcal{A}^{**}(E_{pk}(m_b)) = b]$. Now, we can construct an attacker $\mathcal{A}'$ that wins the IND-CCA1 game as follows: $\mathcal{A}'$ invokes $\mathcal{A}^{**}$ on input of the IND-CCA1 challenge $c_b$, and output whatever $\mathcal{A}^{**}$ guesses. Inequality (2) tells that the result of $\mathcal{A}^{**}$ differs from that of $\mathcal{A}$ (on a syntactically correct input) with a probability of less than $\varepsilon$. Moreover, $\mathcal{A}$ would by assumption guess correctly with an advantage of at least $2\varepsilon$. So by the second triangle inequality, and with the abbreviation $I = (E_{pk}(m_b), g^m h^{r'}, E_{pk}(r'))$, we get

$$
\begin{aligned}
\Big| \Pr[\mathcal{A}^{**}(c_b) = b] - \frac{1}{2} \Big| \\
\geq \Big| |\Pr[\mathcal{A}(I) = b] - 1/2| \\
- |\Pr[\mathcal{A}(I) = b] - \Pr[\mathcal{A}^{**}(c_b) = b]| \Big| \\
\geq 2\varepsilon - \varepsilon = \varepsilon,
\end{aligned}
$$

where $\mathcal{A}^{**}$ has complexity $\tau + q \cdot (t_{ED} + t_{eim}) + t_{ED} - 1$, taking at most $q$ queries, which contradicts the assumed IND-CCA1-security of $E$.

## IV. Blind Turing-Machines

Informally, a blind Turing-machine (BTM) is a normal TM, having its tape alphabet and transition function encrypted under a homomorphic public-key encryption with plaintext equality checking. The transition between states is made by homomorphic manipulations, and the choice of the current transition is made upon plaintext comparisons. We describe the construction over a sequence of subsections to follow.

### A. Definitions

We start with a standard two-way infinite tape Turing-machine $M = (Z, \Sigma, \delta, s)$, working over a tape alphabet $\Sigma$ with $Z$ being its state-space (including the halting state), and $s \in Z$ being the initial state. The mapping $\delta$ describes the state transitions in terms of transforming configurations of the TM to one another. A *configuration* is a tuple $(q, w_L, \sigma, w_R) \in Z \times \Sigma^* \times \Sigma \times \Sigma^* =: Conf_M$, describing the fact that the machine is currently in state $q \in Z$, with symbol $\sigma \in \Sigma$ under its head, and with $w_L, w_R \in \Sigma^*$ being the words to the left- and right of the head. The transition function $\delta: Conf_M \to Conf_M$ is a finite set of transformations $\delta(q, w_L, \sigma, w_R) = (p, w_L', \sigma', w_R')$, i.e., $M$ moves to state $p$ and modifies the tape content toward $w_L', w_R'$ and $\sigma'$. Without loss of generality, we restrict our attention to *deterministic* TM here, as there is no conceptual difference in the nondeterministic case, except that we work on a transition *relation* rather than a function (as will become clear below, the necessary changes to define blind nondeterministic TM are all obvious).

We abbreviate configurations as $\chi$ and write $\chi_1 \vdash \chi_2$ as a shorthand of $\delta(\chi_1) = \chi_2$. A *computation* of $M$ on an initial configuration $\chi_0$ is a finite sequence of configurations $\chi_0 \vdash \chi_1 \vdash \chi_2 \vdash \cdots \vdash \chi_\tau = (h, \dots)$ that ends in a halting state $h \in Z$ and output configuration $\chi_\tau$. The number $\tau$ of steps is called the machine's *time-complexity*, which normally depends on the size of the input (polynomial mostly, if we are after efficient algorithms).

Notice that for our purposes, we do not distinguish moving steps (where only the head is relocated) from substitution steps (in which the current symbol on the tape is replaced by something else). Also note that it is difficult to hide the head movements from the execution environment of the TM (e.g., a universal TM), yet it is necessary to "decorrelate" the head movement pattern from the tape content to achieve confidentiality of the overall computation. Otherwise, the movement of the TM discloses the tape content length and perhaps even reveals the current action that is been carried out (by virtue of some characteristic moving sequences, as would perhaps be the case for pen-and-paper multiplication or division by repeated subtraction which reveals the quotient via counting the number of subtractions, regardless of whether or not they are encrypted).

In section V, we will look at necessary precautions to prevent leakage of information from the Turing-machines head movements alone (quasi as a side-channel to the data as such). Note that similar concerns may apply to garbled circuits as well, as the way in which circuit gates (whether or not they are garbled) are interconnected may already leak partial information about the circuit's potential functionality. Still, we emphasize that our main goal in this work is to protect the data being processed. Hiding the algorithm itself from the execution environment is subject of future considerations and outside the scope of this current work.

### B. Encoding of States and Tape-Symbols

Take a conventional TM $M = (Z, \Sigma, \delta, s)$. Let HPKEET operate on the plaintext space $\mathbb{G}_m$ and ciphertext space $\mathbb{G}_c$, and fix an (invertible) encoding $C: Z \times \Sigma \to \mathbb{G}_m$, so that we can encrypt both, the state *and* current tape symbol.

Computations are usually done over relatively small alphabet, say bits ($\Sigma = \{0,1\}$) or radix-10 numbers ($\Sigma = \{0, \dots, 9\}$). Moreover, the number of states can be expected to be feasibly small as well (otherwise, the representation of $M$ could not be handled by the universal TM in feasible time). Hence, if $|\mathbb{G}_m| = 2^{O(\kappa)}$ for some security parameter $\kappa$, then high min-entropy in the sense of (1) can be assured by sufficiently large $\kappa$ and by assigning random and unique representatives from $\mathbb{G}_m$ to each element of $\Sigma \times Z$, in order to thwart trial decryptions succeeding in polynomial time.

### C. Construction

Our blind TM works over ciphertexts only, and does its transitions using a proper „encryption" of the original state transition function in $M$. To this end, we extend $M$ toward $\widehat{M} = (\mathbb{G}_c, \mathbb{G}_c, \hat{\delta}, \hat{s})$ and define the blind TM as the pair $BTM = (\widehat{M}, \text{HPKEET})$. Here, the $\wedge$-accent is used to denote the „encrypted"counterparts of the respective elements in $M$'s description. We stress that the description is technically complete but to this extent insecure, as the head movement pattern may leak information about the tape content. For the sake of a complete description at this point, however, we postpone the necessary details on security to section V. A blind TM works exactly as a normal TM, but employs HPKEET to do transitions over encrypted configurations as follows:

1) *Encrypted configurations:* given a configuration $\chi = (q, w_L, \sigma, w_R)$ of $M$, the respective encrypted configuration $\hat{\chi}$ *(under the public key $pk$, which we omit in the following to simplify our notation), is defined as*

$$\chi := (\text{Enc}_{pk}(p), w_L', \text{Enc}_{pk}(\sigma), w_R'),$$

Where $w_L', w_R'$ are the encryptions of the tape content under $\text{Enc}_{pk}$ in electronic codebook mode.

2) *Transition functions:* for each pair of consecutive configurations $\chi = (q, w_L, \sigma, w_R) \vdash \chi' = (p, w_L', \sigma, w_R')$ of $M$, the transition function $\hat{\delta}$ for the blind TM is created from $\delta$ as

$$\hat{\delta} = \{\hat{\chi} \mapsto (\text{Enc}_{pk}(\Delta_p), w_L', \text{Enc}_{pk}(\Delta_\sigma), w_R') \mid \delta(\chi) = \chi'\}$$

Where $\Delta_p = p - q$ and $\Delta_\sigma = \sigma' - \sigma$, both computed in $(\mathbb{G}_m, +)$. So, unlike the transition of the TM, a blind TM encrypts only the "difference" between the current and next configuration, in order to enforce re-randomization via homomorphic manipulations on the ciphertexts. Hence, actually doing a transition is now a two-step process:

a) *We invoke* Com *with the token* $t \leftarrow \text{Aut}(sk)$ *on the current (encrypted) configuration* $\hat{\chi}$ *of the BTM to match the states and symbols, and retrieve* $\hat{\delta}$.

b) *We create the new configuration* $\chi'$ *from the current one* $\chi = (Enc_{pk}(q), w_L, Enc_{pk}(\sigma), w_R)$ *by computing in* $(\mathbb{G}_c, \cdot)$,

$$\hat{\chi}' = (\text{Enc}_{pk}(q) \cdot \text{Enc}_{pk}(\Delta_p), w_L', \text{Enc}_{pk}(\sigma) \cdot \text{Enc}_{pk}(\Delta_\sigma), w_R')$$

so as to resemble $M$'s original move via the homomorphic properties of Enc, which is easily verified to be

$$\hat{\chi}' = (\text{Enc}_{pk}(q + \Delta_p), w_L', \text{Enc}_{pk}(\sigma + \Delta_\sigma), w_R')$$
$$= (\text{Enc}_{pk}(p), w_L', \text{Enc}_{pk}(\sigma'), w_R') = \widehat{\chi'}$$

Doing tape manipulations by other means than homomorphic transformations for the sake of stronger IND-CCA2 security is potentially insecure, as we will discuss in a little more detail in section V.A.

Based on this construction, it is a trivial matter to decrypt and recover the tape content by virtue of $\text{Dec}_{sk}$, and we omit the details here. However, note that like in a setting of fully homomorphic encryption, the state transitions require the token as an "evaluation key".

## V. Security Of Blind Computations

By construction, an execution of a BTM produces a sequence of encrypted configurations, enjoying a one-to-one correspondence to the respective sequence of configurations arising from an execution of $M$. However, to retain indistinguishability in experiment $\mathbf{Exp}^{\mathcal{A}}_{\text{IND-CCAE1}}$, we ought to equalize the length of computations on inputs of equal length, *and* make the head movements indistinguishable over different inputs. To this end, we must transform the given Turing-machine accordingly before turning it into a blind TM.

To equalize the length of computations, we restrict the time-complexity bound of $M$ to time-constructible functions. We say that a function $T: \mathbb{N} \to \mathbb{N}$ is *time-constructible*, if there is a Turing-machine $M_T$, which for every input $w$ of length $|w|$ takes exactly $T(|w|)$ steps for its computation on $w$ (not necessarily saying that it computes anything useful). For example, every polynomial function is time-constructible (but also exponential functions, sums, products and compositions of time-constructible functions retain this property).

Furthermore, we must logically decouple the tape content from the head movement pattern to avoid leakage of information via tracking what the head of the blind TM does. Turing-machines whose head movements are a function of the time only (hence independent of the tape content) are called *oblivious* TM. Besides theoretical interest in these for the sake of constructing circuits, the following well-known theorem will help to establish security of blind computations:

**Theorem 1.** (Pippenger and Fischer [22]) Any Turing-machine that runs in time $\tau$ can be simulated by an oblivious Turing-machine in time $O(\tau \log \tau)$.

A naive yet constructive approach to create an oblivious TM from a given one is to mark where the head of the tape is and then scan the tape to locate the head marker in each step. This yields a suboptimal time bound of $O(\tau^2)$ for a running time of $\tau$ on the original TM, and Theorem 1 gives in fact the optimal bound.

So, given a Turing-machine $M$ whose time-complexity is a time-constructible function, we first transform $M$ into an oblivious Turing-machine $M'$, running in time $\tau_M$, which is again time-constructible by some Turing-machine $M_T$. Then, we let our blind TM run $M_T$ in parallel to $M'$ on a second tape, so as to equalize the length of its computation, while running the oblivious TM $M'$ to do the actual computation with head movements that are independent of the data. This proves the following (intermediate nevertheless important) statement:

**Theorem 2.** Let $M$ be a Turing-machine, whose time complexity $T$ is a time-constructible function. Then there exists a functionally equivalent Turing-machine $M'$ with the following property: given any two input words $w_1 \neq w_2$ of the same length $|w_1| = |w_2|$, a computation of $M'$ takes identical head movements on both, $w_1$ and $w_2$.

We can now turn to the task of lifting security assurances that hold for HPKEET towards security for an entire computation on a blind oblivious Turing-machine. Notice that so far, we considered security only for *one* message to be deciphered (as in $\mathbf{Exp}^{\mathcal{A}}_{\text{OW-CCA(E)1}}$) or recognized (from two given ones, as in $\mathbf{Exp}^{\mathcal{A}}_{\text{IND-CCA(E)1}}$). Security of a computation of a blind TM, however, requires a slight change to the experiments, in the sense that the challenge-phase in both games now itself is repeated a number of times that equals the time-complexity[3] $\tau_M$ of the underlying TM $M$. Omitting the obvious details on the changes to the experiments here for brevity, let us directly turn to the respective security conclusions about HPKEET under $\tau_M$ many encryptions (each one of which arises along the emulation of $M$ by a blind TM).

**Lemma 5.** If HPKEET is $(\tau, q, \varepsilon)$-IND-CCA1 secure for a single encryption, then it is $(\tau - n \cdot t_{\text{Enc}}, q, n \cdot \varepsilon)$-IND-CCA1-secure for $n$ encryptions, when $t_{\text{Enc}}$ bounds the complexity of an encryption using HPKEET. Given the additional hypothesis that all random encrypted plaintexts have high min-entropy in the sense of Lemma 1, then the system is also $(\tau - n \cdot t_{\text{Enc}}, q, n \cdot \varepsilon)$-OW-CCA1-secure for $n$ encryptions.

*Proof (sketch).* Indistinguishability is shown by assuming the existence of a pair of (with probability $\geq n \cdot \varepsilon$) distinguishable $n$-tuples, and constructing hybrids to infer distinguishability in the single-message case with probability $\varepsilon$. To this end, the distinguisher must emulate encryptions of no more than $n$ (fixed) input messages, which enlarges the circuits (and yields the modified complexity-bounds).

Onewayness is analyzed in a similar fashion, but is slightly simpler in the details: if a circuit $\mathcal{A}$ exists that upon input of $n$ ciphertexts ouputs one of the underlying plaintexts with probability $\geq n\varepsilon$, then a new circuit $\mathcal{A}'$ can be constructed to correctly answer a single challenge in $\mathbf{Exp}^{\mathcal{A}'}_{\text{OW-CCA(E)1}}$ as follows: $\mathcal{A}'$ randomly constructs $n - 1$ challenges $\tilde{c}^*_1, \dots, \tilde{c}^*_{n-1}$ (adding the complexity $\leq n \cdot t_{\text{Enc}}$), and invokes $\mathcal{A}$ on these challenges along with the given challenge $\tilde{c}_n = c^*$. With probability $\geq n \cdot \varepsilon$, $\mathcal{A}$ outputs one plaintext $\tilde{m}_i$ from the $n$ ciphertexts. The chances for this to be $m^*$ are $\frac{1}{n} n \varepsilon = \varepsilon$, contradicting the presumed OW-CCA1-security of HPKEET.

We stress that onewayness is analyzed under the assumption that plaintext comparisons are possible. Therefore, we must assume high min-entropy of plaintexts, but cannot – and in fact do not – rest on a secret encoding (as described in section IV.B, as this would prevent the constructed machines from emulating proper encryptions (for the same reason, security of multiple encryptions fails in the secret key paradigm). For one-wayness to be assured, however (and fortunately), we do not need the encoding function in the technical arguments, since it is easy to generate random plaintexts whose min-entropy is high, even if these do not lie in the image of the encoding function that the honest party potentially uses. Under the additional min-entropy assumption, trial decryptions under a complexity bound $\tau$ or better ($\leq \tau - n \cdot t_{\text{Enc}}$) are ruled out. □

---

[3] Note that we do not need the space-complexity here, as we only need to count (bound) the number of modifications on the tape, which is bounded by the number of transitions, which is the time-complexity.

Since a BTM basically produces a sequence of ciphertexts rather than a single one, it is a simple matter to instantiate the concrete security parameters of HPKEET based on Lemma 5. Theorem 3 assumes an oblivious Turing-machine to be available, which is assured by theorem Theorem 2.

**Theorem 3.** Let $M$ be an oblivious deterministic Turing-machine with time-complexity $\tau_M$. If HPKEET is $(\tau, q, \varepsilon)$-OW-CCA1/IND-CCA1 secure, then the execution of the BTM constructed from $M$ is $(\tau - \tau_M \cdot t_{\text{Enc}}, q, \tau_M \cdot \varepsilon)$-OW-CCA1 secure. Furthermore, if $\tau_M$ is time-constructible by another Turing-machine $M_\tau$, and if the oblivious BTM emulates (on two parallel tracks on its tape) the executions of both, $M$ and $M_\tau$, stopping not before both executions terminate, then the execution is also $(\tau - \tau_M \cdot t_{\text{Enc}}, q, \tau_M \cdot \varepsilon)$-IND-CCA1 secure.

Note that the apparently awkward mix of time-complexities and circuit complexities that appears in the above statement is actually meaningful, as the time-complexity merely determines how many ciphertexts an execution of an algorithm will provide to the cryptanalytic circuit (being an adversary of type 1). Hence, the circuit complexity is somewhat proportional to the time-complexity.

### A. (In)security of Non-Homomorphic Transitions

Encryptions with equality checks have been designed earlier [19] under the stronger notion of security against adaptive chosen ciphertexts (OW/IND-CCA2), which makes the encryption necessarily no longer homomorphic. Doing a transition by a humble replacement of the current tape ciphertext (encrypted symbol) by another is possible, yet removes the indistinguishability property of computations, because an (encrypted) symbol will always and necessarily be replaced by the *same* symbol, even if the computation itself is different. As a consequence, distinct plaintexts $m_1 \neq m_2$, even if they are equally long, can be distinguished by an external instance upon different sequences of configurations. This can be done without the Aut-token, so that the computation would be insecure in our modified IND-CCA1 game (where the challenge phase includes multiple ciphertexts), and hence be insecure in an adaptive chosen-ciphertext scenario too.

### VI. PUTTING A BLIND TM TO WORK

With the ground prepared in previous sections, we now give the complete picture of how a blind TM is created and envisioned to work in a potentially hostile environment. Let Alice be the honest party who wishes to have her data processed externally by a service provider (SP), having a public key $pk_{SP}$. Alice has her own secret/public key pair ($pk_A$, $sk_A$). For the sake of practicality, let us assume that Alice uses Damgårds Version of ElGamal encryption for ($G, E, D$), which is multiplicatively homomorphic and known to be CCA1-secure [20]. To change the multiplicative homomorphic property into an additive one, Alice encrypts commitments $g^m$ instead of $m$, so that the HPKEET ciphertexts now take the form $(E_{pk_1}(g^m), g^m h^r, E_{pk_2}(r))$. Assuming that the tape alphabet and number of states of her TM is feasibly small, recovering $m$ from $g^m$ is doable via lookup-tables. This adds an additional commit/decommit stage– shown dashed – in Fig. 1, where the overall process is sketched, including locations of type 1 and type 2 adversaries.

To have the SP process her data using a Turing-machine $M$, while not learning anything about it, Alice performs the steps below.

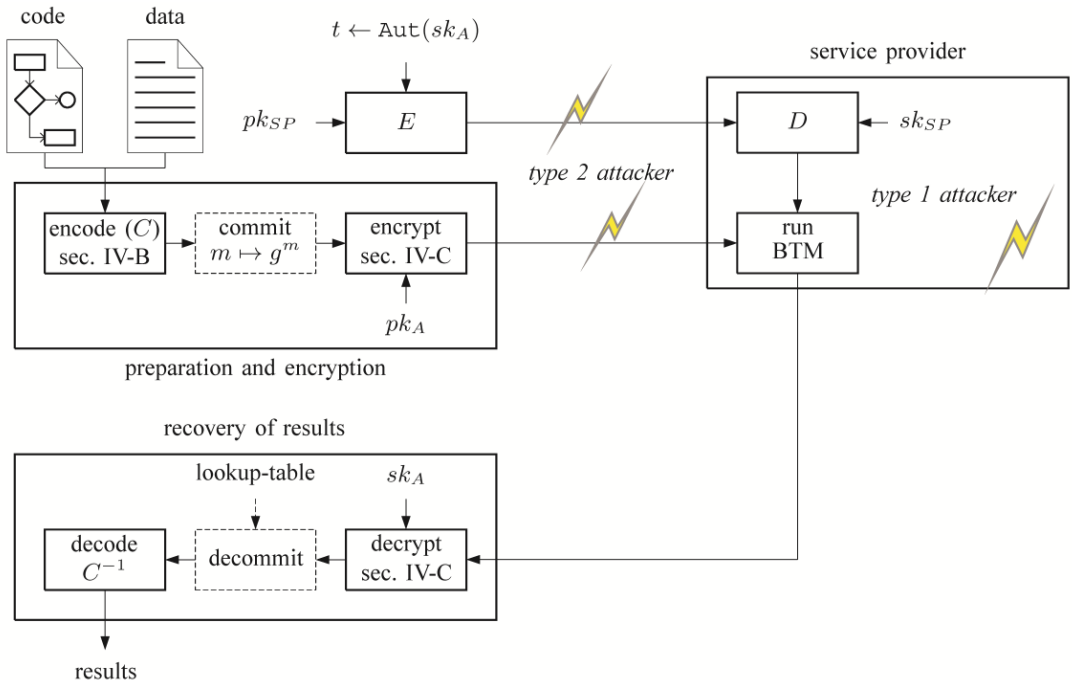

Fig. 1. Usage scheme of a blind Turing machine

*1) She constructs an oblivious Turing-machine $M'$ that simulates $M$ and on a second track/tape (obliviously) runs the machine $M_T$ that takes exactly $\tau_M(|w|)$ steps to terminate for an input $w$. This is to equalize the length of computations and head movements, regardless of the actual input. Call the resulting Turing-machine $M$ (again, for simplicity).*

*2) She constructs a blind Turing-machine $\hat{M}$ (code) from $M$ as described in section IV. In doing so, she prepares the tape content (data) in a three steps:*

*a) Encode each tape symbol and state by the function $C$ (to assure high min-entropy).*

*b) Compute a commitment to each encoded state and symbol (to make the multiplicatively homomorphic encryption additively homomorphic; this step can be omitted if $(G, E, D)$ is additively homomorphic already, hence is shown dashed in Fig. 1).*

*c) Encrypt the commitment under the public key $pk_A$.*

*3) She then sends all information to the service provider, potentially under the eyes of a type 2 attacker (cf. section II), against which Theorem 3 assures IND-CCA1 security.*

*4) She submits the authorization token $t \leftarrow \text{Aut}(sk_A)$ in encrypted form (under the public key $pk_{SP}$) to the service provider. Observe that the encrypted authorization token plays the role of something like a "license" to execute the given Turing-program, which would otherwise not be possible.*

The service provider executes the (encrypted) code, i.e., runs the blind Turing-machine on the encrypted tape content, and returns the encrypted tape content. While doing so, the SP may attempt to learn information from the execution of the BTM or the intermediate tape contents, in which case the SP becomes a type 1 attacker (cf. section II), against which Theorem 3 assures OW-CCA1 security.

The decryption of the ouput tape content is done as the encryption, only in reverse order, and by virtue of lookup tables to „decommit" the decrypted commitments $g^m$, if there has been a commitment stage during the data preparation. The results are finally available after decoding (function $C^{-1}$).

TABLE I. COMPLEXITIES (IMPLEMENTATION-RELATED)

| Object/Action | | Complexity |
|---|---|---|
| **HPKEET** | KeyGen | $2 \times G$ |
| | Enc($m$) | $2 \times E + 2e + m$ |
| | $g^m \leftarrow$ Dec | $2 \times D + e + m$ |
| | Aut | $O(1)$ |
| | Com | $2(D + e + i + m$ |
| **blind TM** | transition selection | $D + O(1)$ |
| | Tape manipulation | $3m$ |

It is as well imaginable to let the program come from a different instance (entity in the system) as the data, given that both instances have agreed on a common encoding. This scenario would be, for example, useful when a software is provided by some vendor $V$, and shall be executed on data that the customer $A$ owns, while protecting the intellectual property of the software vendor. If the execution of the software shall be outsourced to an SP, then $V$ and $A$ both submit their authorization tokens to the SP, while $A$ and $V$ agree on some common encoding to have the data compatible with the code. The SP then acts as usual to compute the results in privacy. The customer $A$ can in that case only receive and decrypt the results, while being itself unable to execute the program as $A$ lacks $V$'s authorization (token).

Another variation could be $A$ encrypting the data under someone else's public key, to make the results available to another (third or fourth) party, which sees neither the input data nor the code.

## VII. PERFORMANCE AND PRACTICAL ASPECTS

Assuming that the universal blind TM can select the proper transition based on the comparison facility of HPKEET, there would be no change in the asymptotic complexity of any function, whether it is computed on a conventional or blind Turing-Machine. More concretely, however, if a function $f: \{0,1\}^* \rightarrow \{0,1\}^*$ on a value $x \in \{0,1\}^n$ can be computed in time $T(n)$, then a blind TM can compute the same result in time $\leq \tau \cdot T(n)$, where $\tau$ is a constant time bound needed to manipulate a state (via homomorphic operations on the ciphertext). A similar argument can be made for the change in the space-complexity, since tape symbols are now encoded as group elements, thus multiplying the required space as well by a concrete constant factor.

Practically, a blind TM will need more time to complete its computation than a conventional TM since it has to find the proper transition based on invocations of Com. However, this can easily be accelerated if the selection is done by a hash-table taking the commitment $g^m$ of the current tape symbol (being a HPKEET ciphertext $(E_{pk_1}(m), g^m h^r, E_{pk_2}(r))$) as the key for hashing. The transition can then be obtained from the hash-table in roughly $O(1)$ steps, as the commitments can reasonably well be assumed as being uniformly distributed (hence ideal for hashing).

Table I shows an overview of the actions involved when computing on ciphertexts, including actions that refer to HPKEET alone, taking into account that tape symbol commitments are encrypted, decrypted and compared (with the obvious changes to the formal descriptions given in section III). Here, the symbols e, i and m stand for *exponentiation*, *inversion* and *multiplication* inside the group $\mathbb{G}_c$. The notation "$k \times X$" where $X \in \{G, E, D\}$ refers to $k$ executions of the respective algorithm implementation of the encryption $(G, E, D)$ underlying HPKEET.

A (not very much optimized) Java implementation of our HPKEET cryptosystem based on Damgård's version of ElGamal encryption brought up some runtime estimates on a 3.6 GHz computer with 8 GB RAM and 64 Bit Windows 7, as shown in Table II. The numbers are based on an average of 100 invocations of Enc, Dec and Com for key lengths of $\kappa \in \{256, 512, 2048\}$ bit (according to current recommendations of the NIST and other bodies). The value for a transition selection and tape manipulations give a rough estimate on how much slower a blind TM will run compared to a conventional TM (i.e., the factor $\tau$ from the first paragraph).

## VIII. OUTLOOK AND OPEN PROBLEMS

A practical topic of future work is the implementation of the concept within a practical computer architecture including assembler code and hardware. Challenges in such a practical implementation may concern the realization of other arithmetic operations such as integer divisions with remainder or logical manipulations. Results on this will be reported in companion and subsequent work.

TABLE II.  BENCHMARK RESULTS (FOR $(G, E, D)$ BEING DAMGÅRD-ELGAMAL ENCRYPTION)

| | Key size | 256 bit | 512 bit | 2048 bit |
|---|---|---|---|---|
| **Running time [ms]** | Enc | 0.93 | 4.07 | 174.52 |
| | Dec | 0.32 | 2.35 | 108.84 |
| | Com | 1.08 | 3.27 | 130.11 |
| | BTM transition selection | 0.16 | 1.25 | 43.64 |
| | BTM tape manipulation | < 0.05 | < 0.05 | 0.16 |

The central contribution here is the insight that (only) additively homomorphic encryption can be used to construct Turing-machines that work on encrypted information only, by virtue of public-key encryption with equality check. Hence, this work adds a fourth alternative to existing approaches to secure function evaluation, besides fully homomorphic encryption, garbled circuits or secure multiparty computation. Unfortunately, the necessary ingredient of additively homomorphic encryption that is secure against chosen ciphertext attacks is surprisingly rare, while non-homomorphic encryptions under stronger security notions are better known. Taking a closer look at why we require homomorphy to do state transitions reveals that the weaker requirement of re-randomization of ciphertexts is actually sufficient to invalidate the arguments of section V.A. An interesting open problem is thus finding encryptions that allow re-randomization of ciphertexts but are still CCA2-secure (if such encryptions exist).